\documentclass[prc,aps,preprint,showpacs]{revtex4}
\usepackage{graphicx}

\begin{document}

\title{Deformed Halo of $^{29}_{9}$F$_{20}$} 

\author{Ikuko Hamamoto$^{1,2}$} 

\affiliation{
$^{1}$ {\it Division of Mathematical Physics, Lund Institute of Technology 
at the University of Lund, Lund, Sweden} \\    
$^{2}$ {\it Center for Nuclear Study, University of Tokyo, Bunkyo-ku, 
Tokyo 113-0033, Japan}}

\begin{abstract}
Using a simple model based on the knowledge of spherical and deformed Woods-Saxon potentials,  
it is shown that the recent observation of halo phenomena 
in $^{29}$F
can be interpreted as evidence for the prolate deformation of the ground state 
of $^{29}$F.  
The prolate deformation is the result of the shell structure, which is 
unique in one-neutron resonant levels, in particular near degeneracy of the  neutron 1$f_{7/2}$ and 2$p_{3/2}$ resonant levels, together with the strong preference of prolate shape by the proton number $Z$ = 9.   
On the other hand, in oxygen isotopes spherical shape is so much favored 
by the proton number $Z$ = 8 that the presence of possible neutron 
shell-structure may not make the system deformed.   
Thus, the strong preference of 
particular shape by the proton numbers 8 and 9, respectively, together with a  
considerable amount of the energy difference between the neutron $1d_{3/2}$ 
and $2s_{1/2}$ orbits in oxygen isotopes seems to play an important role 
in the phenomena of oxygen neutron drip line anomaly, as was suggested by H.Sakurai {\it et al.} in 1999.   

\end{abstract}

%\begin{keyword} 
%deformed halo \sep shell-structure of weakly-bound neutrons \sep prolate $deformation of F-isotopes\sep oxygen neutron drip line anomaly}
%\end{keyword}

%\newpageafter{abstract}

\maketitle

\newpage

\section{INTRODUCTION} 
While the heaviest bound oxygen isotope that has the magic proton-number 
$Z$=8 is known as $^{24}_{8}$O$_{16}$, the presence of bound fluorine nuclei, 
 $^{29}_{9}$F$_{20}$ and $^{31}_{9}$F$_{22}$, is known.  Namely, while the heaviest bound isotope of oxygen accommodates only up to 16 neutrons, 
that of fluorine ($Z$=9), which has only one proton more, can accommodates 
up to 22 neurons.  
Though having one proton more in fluorine isotopes makes the neutron 
potential stronger so as to accommodate more neutrons, 
the increase of 6 (= 22-16) neutrons in the heaviest bound  
fluorine isotope is unusual in the region of light nuclei.         
For example, in Ref. \cite{DSA19}, the heaviest bound isotopes of fluorine and neon are experimentally confirmed to be $^{31}_{9}$F$_{22}$ and $^{34}_{10}$Ne$_{24}$, respectively.   Namely, the increase of the neutron number in the heaviest bound isotope of neon ($Z$=10) from that of fluorine 
is only 2.   The physics involved in the increase of 
6 neutrons, sometimes called ''oxygen neutron drip line anomaly'', has been  discussed in the literature.    
In the abstract of Ref. \cite{HS99}, in which evidence for particle 
stability of $^{31}$F was reported, it is stated that oxygen 
neutron drip line anomaly may demonstrate the onset of deformation for neutron-rich fluorine isotopes.   
The only direct experimental indication of the deformation of 
the ground state of $^{29}$F, which is known before 2020, is the result of the mass measurement of $^{29}$F reported in Ref. \cite{LG12}.   
In a recent article \cite{SB20} it is reported that the nucleus $^{29}$F shows a two-neutron halo by measuring the reaction cross section.  In the present paper I first concentrate on the interpretation 
of the recently observed halo phenomena in $^{29}$F.   

In order to study the halo phenomena, I try to use 
a simple model based on the knowledge of spherical 
and deformed Woods-Saxon potentials.   This is  
because nuclear shape is anyway one-body property of the mean field, 
and the very basic and simplest description of many-body systems is a  
(self-consistent) mean-field approximation to the many-body problem.   
It is important to note that in deformed nuclei one-particle picture 
is known to work \cite{BM75} much better than in spherical nuclei because 
the major part of the residual interaction in spherical mean field, 
quadrupole-quadrupole force, is already included in the deformed mean field.  
The quantitative success in the analysis of the data especially in some light odd-A nuclei such as $^{25}_{12}$Mg$_{13}$ by using the one-particle picture in the deformed mean field can be found, for example, in Ref. \cite{BM75}.  
In particular, in order to pin down the appearance 
of halo phenomena, it is absolutely necessary to use the models in which 
radial wave functions of weakly-bound particles are properly treated.      
In order to obtain such wave functions of deformed halo nuclei with halo radial dependence, the simple model used in the present paper 
seems to be more appropriate than other model calculations available 
in the market. 

Taking some simple one-body potentials the relation between the preferred shape 
and the degree of shell filling can be analytically studied.   
In a single $j$-shell oblate (prolate) shape 
is favored in the beginning (at the end) of the shell filling, while in the harmonic-oscillator potential prolate (oblate) shape is favored in the beginning of (at the end of) the filling of a major shell.   
The characteristic feature of the shell structure expressed by the harmonic oscillator potential is modified in realistic potentials decisively 
by the presence 
of the spin-orbit potential as well as the feature of finite-well potentials.  
Nevertheless, it is found that some important gross feature except some  phenomena such as halo can be seen to remain 
in simple but slightly more realistic one-body potentials, for example, in a modified oscillator potential \cite{BM75}. 

The shape of a given nucleus is determined by the shell structure 
of both protons and neutrons.  The reason why fluorine isotopes can survive 
up to $N$=22 might be guessed because the proton number $Z = 9$ strongly  favors prolate deformation, as seen from the strong deformation-dependence of the energy of  
the [220 1/2] orbit, which the 9th proton should occupy.   
The dependence can be seen, for example, in Fig. 5-1 of Ref. \cite{BM75}.    
It is noted that for prolate deformation 
the [220 1/2] orbit is the lowest-lying orbit among the orbits coming from  
the $sd$ major shell.     
However, it has been so far difficult to establish experimentally evidence for the deformation of those fluorine isotopes close to the neutron drip line. 
Besides the mass measurement of $^{29}$F in \cite{LG12}, it is reported 
in Ref. \cite{PD17} that the only bound excited state of $^{29}$F is 
observed at $E_{x}$ = 1080 (18) keV.   This rather low excitation energy may be certainly interpreted as an indication of the deformed ground state 
in $^{29}$F.   
The spin assignment of the ground and first excited states of  $^{29}$F in Ref. \cite{PD17} 
is based on shell-model calculations and is so far not experimentally fixed.
On the other hand, studying neutron removal from $^{29}$F beams, 
in Ref. \cite{AR20} it is proposed that the ground state of $^{28}_{9}$F$_{19}$
has $p$-wave valence neutron configuration.  This may indicate that $^{28}$F, 
the neighbor nucleus of $^{29}$F, is deformed.    

Immediately after Ref. \cite{HS99} the structure of neutron-rich F isotopes 
was studied in detail by means of large-scale shell-model calculations 
\cite{UT01}.   The calculated result of $^{29}$F in Ref. \cite{UT01}, 
in particular, shows large probabilities of 2p-2h and 4p-4h components in the wave-function indicating the deformation of the ground state of $^{29}$F.   However, harmonic-oscillator wave-functions used in the shell-model calculation and np-nh expansions make impossible to pin down properly the properties depending on halo of deformed nuclei.      

In the present paper I show that the nucleus $^{29}$F is presumably prolately deformed because, then, 
the presence of the neutron halo phenomena can be very naturally understood.   
In other words, the observation of halo phenomena in the nucleus $^{29}$F 
can be taken as evidence for prolate deformation of 
the ground state of $^{29}$F.  
The proton number $Z$ = 9 
in fluorine may definitely favor the prolately deformed bound system and, consequently, the bound isotopes of fluorine with $N$ = 20 and 22 may be present. 

Deformed (neutron) halo expresses the halo phenomena to be observed in some deformed nuclei close to the (neutron) drip line.   Neutron shell structure and the resulting possible deformation in neutron-drip-line nuclei are discussed 
in Refs. \cite{IH07, IH10, IH12}.   The change of the shell structure in weakly-bound neutrons is described in p.238-240 of Ref.\cite{BM69}.   
Deformed halo has been indeed observed in several odd-N nuclei: for example, 
in $^{31}$Ne \cite{TN09, TN14}, in $^{37}$Mg \cite{NK14} and in $^{29}$Ne 
\cite{NK16}.   The degeneracy of the $f_{7/2}$- and $p_{3/2}$-shells and the consequence of Jahn-Teller effect \cite{JT37} play a crucial role in the formation 
of deformed halo in these odd-N nuclei as well as in the present nucleus $^{29}$F.   

In Sec. II the main points of the model used are briefly summarized,   
and the neutron one-particle levels as a function of axially-symmetric quadrupole deformation 
are shown for the standard Woods-Saxon potential, of which parameters are adjusted for $^{29}$F.  
Examining the figure obtained one finds that in the $N = 20$ system with weakly-bound neutrons  
the prolate deformation of 
$\beta \approx 0.4$ may be energetically favored as much as spherical system.
On the other hand, since the proton number $Z$ = 9 would absolutely favor 
prolate deformation, it may be expected that the nucleus 
$^{29}$F may be prolately deformed.   Then, the structure of the 
possible deformed halo for the prolate deformation will be examined.           
Conclusion and discussions are given in Sec. III.

\section{MODEL AND NUMERICAL CALCULATIONS}
In Fig. 1 I show the neutron one-particle energies as a function of axially-symmetric quadrupole deformation.   Negative energies are eigenvalues 
of the deformed Woods-Saxon potential, while positive energies express one-particle resonant energies which are defined in terms of the eigenphase 
$\delta_{\Omega}$ \cite{IH05}.  Some neutron numbers, which are obtained by filling all lower-lying levels, are indicated with open circles.   

Since the model used is explained in detail in Ref. \cite{IH19}, here I summarize only the important points.  
As a spherical one-body potential which is an approximation to the (self consistent) mean field, I take the phenomenological Woods-Saxon potential 
described in p.238-240 of Ref. \cite{BM69}.  Then, I examine the shell
structure of one-particle energy spectra of neutrons 
in the axially-symmetric quadrupole-deformed potential which is constructed  
in a standard way.  
I write the single-particle wave-function in the body-fixed (intrinsic) 
coordinate system as 
\begin{equation}
\Psi_{\Omega}(\vec{r}) = \frac{1}{r} \, \sum_{\ell j} R_{\ell j  \Omega}(r) \,  
{\bf Y}_{\ell j \Omega}(\hat{r}),
\label{eq:twf}
\end{equation}
which satisfies 
\begin{equation}
H \, \Psi_{\Omega} = \varepsilon_{\Omega} \, \Psi_{\Omega}   
\label{eq:sheq}
\end{equation} 
where $\Omega$ expresses the component of one-particle angular-momentum 
along the symmetry axis and is a good quantum number.   
The values of the parameters in the potential 
are taken from Ref. \cite{BM69}.   
For example, the diffuseness $a$ = 0.67 fm and 
the radius $R$ = $r_{0} A^{1/3}$ with 
$r_0$ = 1.27 fm.   
The depth of the potential depends on the mass number $A$, the proton number 
$Z$ and the neutron number $N$.   For example, if one proton is added to a system, the depth of the neutron potential becomes deeper reflecting the strong attractive neutron-proton interaction.     
    
The coupled differential equations for the radial wave 
functions of one-particle orbits $R_{\ell j  \Omega}(r)$, 
which are obtained from 
the Schr\"{o}dinger equation, are integrated in coordinate space with the correct asymptotic behavior in respective $(\ell,j)$ channels for 
$r \rightarrow \infty$.  
In this way, the $r$ dependence of  
low $(\ell,j)$ components of weakly-bound low-$\ell$ neutrons 
expresses properly possible halo effect.   See, for example, Fig. 2, in which the $r$-dependence of various components of 
the weakly-bound neutron orbit [330 1/2] that is the only $\Omega^{\pi}$ = 1/2$^{-}$ orbit expressed by a long-dashed curve in Fig. 1, is shown.   
Indeed, in Fig. 2 
it is seen that $\ell = 1$ components exhibit clear halo phenomena, while 
the $r$-dependence of $\ell = 3$ components is not so much different from that of well-bound case.  For comparison, see Fig. 6 in Ref. \cite{IH19} for a well-bound case of the [330 1/2] orbit.   
In Fig. 2 one sees a typical example of the $r$-dependence of radial 
wave functions of various $\ell$ components of weakly-bound neutrons 
in the case of deformed halo.    

For spherical shape ($\beta$ = 0) in Fig. 1 the large energy gap at $N$ = 20 
is kept, namely one may say that the magic number $N$ = 20 remains 
for spherical shape.  On the other hand, the energy of the neutron 
Fermi level of 
the $N$ = 20 system is nearly equal for $\beta$ = 0 and 
$\beta$ $\approx$ 0.44.     Noting that the shape of a nucleus is determined by both protons and neutrons and, furthermore, 
the $Z$ = 9th proton outside the magic number $Z = 8$ 
favors prolate deformation, from Fig. 1 
it may be natural   
to expect that 
the prolately-deformed state of $^{29}$F is lower in energy than the spherical state.  Namely, 
the ground state of $^{29}$F is not spherical 
but presumably prolately deformed with $\beta \approx 0.4$.    
Now, if the ground state of $^{29}$F has such prolate deformation,  
the neutron negative-parity level with $\Omega^{\pi} = 1/2^{-}$,   
which comes from the $pf$ major shell and is often called [330 1/2], is just occupied.          

I note that the reason why in Fig. 1 the [330 1/2] level, which is the lowest-lying level among the levels coming from the $pf$-shell, comes down so steeply as $\beta(>0)$ increases from zero is that in the spherical shape 
($\beta = 0$) the resonant $2p_{3/2}$ level lies energetically close to and 
even lower than the resonant $1f_{7/2}$ level.   Consequently, as the deformation sets in, the $2p_{3/2}$ and $1f_{7/2}$ levels immediately mix strongly.   In other words, the possible prolate deformation of $^{29}$F
comes from the shell structure unique in one-particle resonant neutrons    \cite{IH16}.   

In Fig. 2 the components of the radial wave function 
$R_{\ell j \Omega=1/2}(r)$ of the neutron [330 1/2] orbit at $\beta$ = 0.38  
are shown as a function of radial coordinate.
It is seen that the occupation of the weakly-bound [330 1/2] orbit exhibits 
halo phenomena due to the components of weakly-bound $p$ neutrons, while 
no weakly-bound negative-parity orbits are occupied in the $N = 20$ system around spherical shape as seen from Fig. 1.   
In short, it can be said 
that the observed halo phenomena in $^{29}$F indicate   
the prolate deformation of the nucleus $^{29}$F.     

One may wonder whether or not there is a considerable correlation 
in the two weakly-bound neutrons.  
In the present paper I do not go into the correlation problem.   
However, around $\beta$ = 0.4 in Fig. 1 
one finds that just below $N$ = 20 the neutron 
$\Omega^{\pi}$ = 1/2$^{-}$ and 1/2$^{+}$ levels are present 
while just above $N$ = 20 
the neutron $\Omega^{\pi}$ = 3/2$^{-}$ and 3/2$^{+}$ levels are present. 
Since those negative-parity weakly-bound neutron levels, $\Omega^{\pi}$ = 1/2$^{-}$ and $\Omega^{\pi}$ = 3/2$^{-}$, are expected to be 
the considerable components of the wave function of 
the possibly correlated two neutrons 
and contain considerable amounts of $2p_{3/2}$ component, 
the appearance of halo phenomena is expected irrespective of the presence or absence of a considerable amount of two-neutron correlation.

\section{CONCLUSION AND DISCUSSIONS} 
Inspired by the recent experimental work reporting the halo phenomena in $^{29}$F \cite{SB20}, I have tried to understand the origin of the halo,    
using one-particle neutron spectra obtained from a standard Woods-Saxon potential.  
The appearance of the neutron halo can be naturally understood if 
the nucleus $^{29}$F is prolately deformed.   The prolate deformation is expected from combining the proton number $Z = 9$ with the one-particle energy spectra of weakly-bound neutrons around $N = 20$ as a function of deformation.   In short, the presence of 
the neutron halo indicates the prolate deformation of the nucleus $^{29}$F.   Furthermore, noting that the oxygen isotopes ($Z = 8$)  strongly favor spherical shape and one expects a considerable energy difference 
 between the neutron $1d_{3/2}$ and $2s_{1/2}$ orbits for spherical shape, 
the shape difference between 
the isotopes of oxygen and fluorine may be an important element in the understanding of 
so-called 
oxygen neutron drip line anomaly, as was suggested in Ref. \cite{HS99}.  
I note that the large energy difference between the neutron $1d_{3/2}$ and $2s_{1/2}$ levels in oxygen isotopes 
has been already recognized, for example, from the fact that the nucleus  
$^{24}_{8}$O$_{16}$ has been sometimes called a doubly magic nucleus.      
For example, see Ref. \cite{RVFJ09}.   

In some available publications, for example Refs. \cite{RVFJ09} and 
\cite{TIT20}, it is suggested that the oxygen neutron drip line anomaly comes 
from the fact that due to the presence of a $1d_{5/2}$ proton in fluorine isotopes the proton-neutron tensor interaction lowers the energy of the neutron $1d_{3/2}$ orbit so as to make a long fluorine neutron drip line.  However, 
this suggestion may explain the drip line extended to $N$ = 20 but not to 
$N$ = 22, $^{31}$F.    Furthermore, a question is whether the strength of the tensor force, which is needed to extend the neutron drip line,   is acceptable or not.    

From Fig. 1 it is seen that the nucleus $^{31}$F can be 
also prolately deformed due to the proton number $Z = 9$, however, 
whether or not 
it shows neutron halo phenomena depends in a subtle way on the size of the deformation, as seen from Fig. 1.    
If the size of the prolate deformation is such that the last two weakly-bound 
 neutrons in $^{31}$F occupy mainly the $\Omega^{\pi}$ = 3/2$^{+}$ orbit 
 coming from 
the $1d_{3/2}$ orbit, the halo phenomena could be rather weak, 
even if the neutron $\Omega^{\pi} = 1/2^{-}$ level is occupied.   

What is described in the present article is based 
on the simple phenomenological one-body potential, Woods-Saxon potential.   Though simple models can be expected 
to show some essential points of physics in a simple way, 
it is hoped that advanced microscopic models, which can explain also halo phenomena of $^{29}$F in an acceptable way, will quantitatively 
show the physics that is described in the present article.

\newpage

\noindent
{\bf\large Figure captions}\\
\begin{description}
\item[{\rm Figure 1 :}] 
Calculated neutron one-particle energies in the potential produced 
by $^{29}$F as a function of quadrupole deformation 
parameter $\beta$.
Some neutron numbers, which are obtained by filling all lower-lying levels, 
are indicated with open circles.   One-particle bound and resonant energies 
at $\beta = 0$ are $-$5.956, $-$3.713, $-$1.041, and $+$3.002 MeV 
for the $1d_{5/2}$, $2s_{1/2}$, $1d_{3/2}$, and $1f_{7/2}$ levels, 
respectively.   
The width of one-particle resonant level of $1f_{7/2}$ is 0.422 MeV. 
The one-particle resonant level of $2p_{3/2}$, which is expected 
below that of $1f_{7/2}$ at $\beta = 0$, is not obtained 
when the one-particle resonance 
is defined in terms of eigenphase. 
The unusual behavior of the $\Omega^{\pi} = 1/2^{-}$ curve 
in the region of $0.1 < \beta  < 0.25$, which is connected to $1f_{7/2}$ 
at $\beta = 0$, is due to the fact that the one-particle resonant level cannot be obtained for 
$\varepsilon_{\Omega} >$ 1 MeV when the major component is $\ell = 1$.   
Nevertheless, figure indicates that at $\beta = 0$ the $p_{3/2}$ one-neutron resonant level lies lower than the $f_{7/2}$ one-neutron resonant level.  
Some one-particle resonant levels ($\varepsilon_{\Omega} > 0$) 
for $\beta \neq 0$, which are defined 
in terms of the eigenphase $\delta_{\Omega}$ 
\cite{IH05},  
are not plotted if they are not relevant to the present interest. 

\end{description}

\begin{description}
\item[{\rm Figure 2 :}]
An example of deformed halo wave-functions: Components of the radial wave function $R_{\ell j \Omega=1/2}(r)$ 
of the neutron $\Omega^{\pi} = 1/2^{-}$ orbit, so-called [330 1/2], 
expressed by a long-dashed curve in Fig. 1,  
at $\beta$ = 0.38 are shown as a function of radial coordinate.  
Respective $(\ell j)$ quantum numbers 
are denoted without writing the radial node quantum-number $n$, 
because the wave functions are not eigenfunctions 
of any spherical potential.  The probabilities of respective $(\ell j)$ 
components in the wave function are 0.070, 0.537, 0.009, and 0.376 for  
$p_{1/2}$, $p_{3/2}$, $f_{5/2}$, and $f_{7/2}$, respectively.  
Since the probabilities of 
$h_{9/2}$ and $h_{11/2}$ are only 0.000 and 0.008, respectively, 
the $h_{9/2}$ 
and $h_{11/2}$ wave functions are not plotted.   
The thin vertical line at $r$= 3.90 fm denotes the radius 
of the Woods-Saxon potential used.
\end{description}


\vspace{2cm}
\begin{thebibliography}{99}
\bibitem{DSA19} D. S. Ahn {\it et al.}, Phys. Rev. Lett. {\bf 123} (2019)  212501.     
\bibitem{HS99} H. Sakurai {\it et al.}, Phys. Lett. B {\bf 448} (1999) 180.  
\bibitem{LG12} L. Gaudefroy {\it et al.}, Phys. Rev. Lett. {\bf 109} (2012) 202503. 
\bibitem{SB20} S. Bagchi {\it et al.}, Phys. Rev. Lett. {\bf 124} (2020) 222504.
\bibitem{BM75} A. Bohr and B. R. Mottelson, {\it Nuclear Structure\/} 
(Benjamin, Reading, MA, 1975), Vol. II.  
\bibitem{PD17} P. Doornenbal {\it et al.}, Phys. Rev. C {\bf 95} (2017) 041301(R). 
\bibitem{AR20} A. Revel {\it et al.}, Phys. Rev. Lett. {\bf 124} (2020) 152502.  \bibitem{UT01} Yutaka Utsuno, Takaharu Otsuka, Takahiro Mizusaki and Michio Honma,  Phys. Rev. C {\bf 64} (2001) 011301(R). 
\bibitem{IH07} I. Hamamoto, Phys. Rev. C {\bf 76} (2007) 054319.   
\bibitem{IH10} I. Hamamoto, Phys. Rev. C {\bf 81} (2010) 021304.   
\bibitem{IH12} I. Hamamoto, Phys. Rev. C {\bf 85} (2012) 064329.
\bibitem{BM69} A. Bohr and B. R. Mottelson, {\it Nuclear Structure\/} 
(Benjamin, Reading, MA, 1969), Vol. I.   
\bibitem{TN09} T. Nakamura {\it et al.}, Phys. Rev. Lett. {\bf 103} (2009) 262501.   
\bibitem{TN14} T. Nakamura {\it et al.}, Phys. Rev. Lett. {\bf 112} (2014) 142501.
\bibitem{NK14} N. Kobayashi {\it et al.}, Phys. Rev. Lett. {\bf 112} (2014) 242501.      
\bibitem{NK16} N. Kobayashi {\it et al.}, Phys. Rev. C {\bf 93} (2016) 014613.  
\bibitem{JT37} H. A. Jahn and E. Teller, Proc. R. Soc. {\bf A161} (1937) 220.    \bibitem{IH05} I. Hamamoto, Phys. Rev. C {\bf 72} (2005) 024301.   
\bibitem{IH19} I. Hamamoto, Phys. Rev. C {\bf 100} (2019) 014324.   
\bibitem{IH16} I. Hamamoto, Phys. Rev. C {\bf 93} (2016) 054328.   
\bibitem{RVFJ09} R. V. F. Janssens, Nature {\bf 459} (2009) 1069.     
\bibitem{TIT20} T. I. Tang {\it et al.}, Phys. Rev. Lett. {\bf 124} (2020) 212502. 


\end{thebibliography}
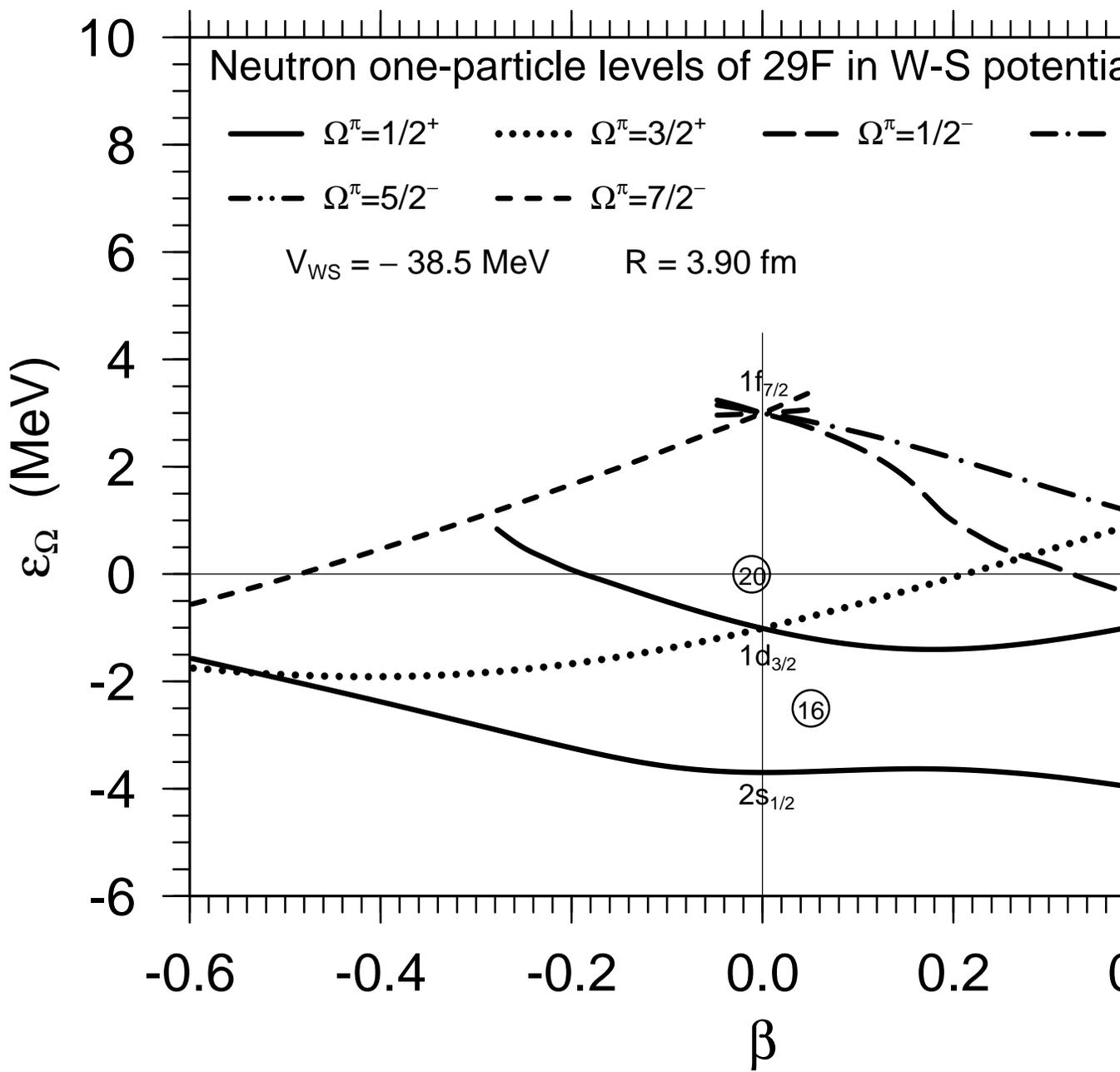
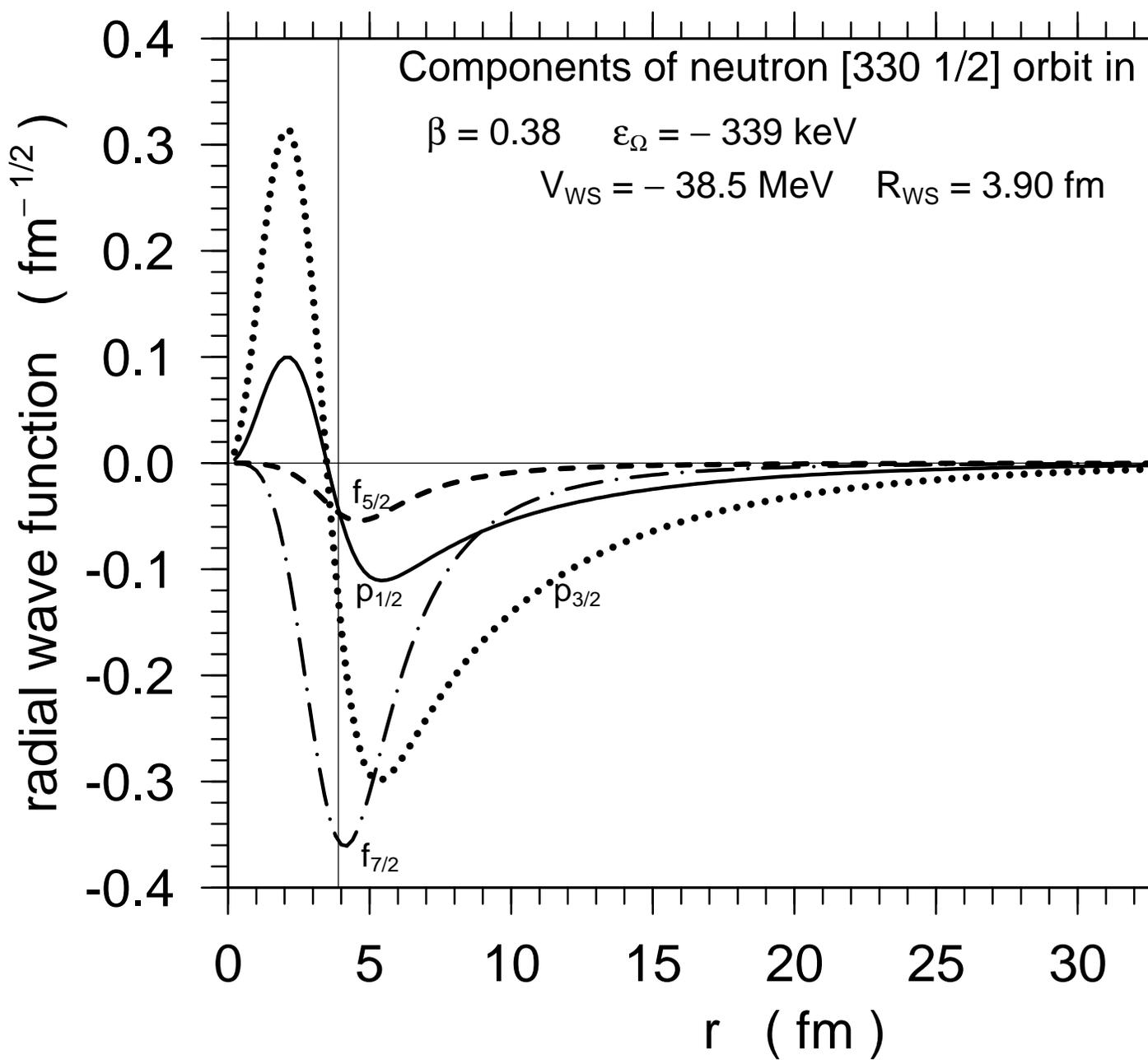
\end{document}